\documentclass[12pt, onehalfspacing, onecolumn]{article}
\usepackage{graphicx, amsmath, amssymb}
\usepackage{psfrag}
\usepackage{epstopdf}
\usepackage{authblk}
\usepackage{float}
\usepackage{hyperref}
\usepackage[a4paper,left=30mm,right=30mm,top=20mm,bottom=15mm,bindingoffset=0cm]{geometry}

\title{Combined experimental/theoretical approach to residual stresses within multiplicative elasto-plasticity }

\author[1, 2]{I. I. Tagiltsev\thanks{i.i.tagiltsev@gmail.com}}
\author[1, 2]{A. V. Shutov\thanks{alexey.v.shutov@gmail.com}}
\affil[1]{Lavrentyev Institute of Hydrodynamics, Pr. Lavrentyeva 15, 630090, Novosibirsk, Russia}
\affil[2]{Novosibirsk State University, Ul. Pirogova 1, 630090, Novosibirsk, Russia}

\begin{document}
\maketitle

%_____________________________________________________________________

\begin{abstract}
The study is devoted to geometrically non-linear modelling of viscoplastic structures with residual stresses.
We advocate and develop a special approach to residual stresses based on the transition between reference configurations.
 The finite strain kinematics of the viscoplastic material is modelled by the multiplicative decomposition of the deformation gradient tensor.
%It is shown that the advocated approach is especially efficient in that case due to the weak invariance of the material model.
 Numerical algorithms originally developed for unstressed materials are extended to materials with pre-stresses.
 Owing to the weak invariance of constitutive equations, the incorporation of pre-stresses happens without additional costs. Thus, the advocated approach is especially efficient.
A novel experimental/theoretical method for assessment of residual stresses in welded structures is presented; the method combines advantages of purely experimental and theoretical approaches.
To demonstrate the applicability of the proposed procedure, we simulate plate welding.
As an example we show that the procedure allows to extrapolate the filed of residual stresses away from the measurement points.
As another example, we address the reduction of weldment-related residual stresses by mechanical treatment.
\end{abstract}

\section*{Nomenculature}

\begin{table}[H]
\begin{tabular}{rl}
%\hline

% model constutients

$\mathbf{F}$ & deformation gradient tensor\\
$\mathbf{C}$ & right Cauchy-Green tensor \\
$\Psi$ & Helmholtz  free-energy per unit mass \\
$\mathbf{\tilde{T}}$ & 2nd Piola-Kirchhoff stress tensor \\
$\mathbf{T}$ & Cauchy stress tensor \\
$s$ & plastic arc-length \\

% essential tensor math

$\mathbf{A}^{\text{T}}$ & transpose of a tensor \\
$\text{tr}(\mathbf{A})$ & trace of a tensor \\
$\mathbf{\overline{A}}$ & unimodular part of a tensor \\
$\mathbf{A}^{\text{D}}$ & deviatoric part of a tensor \\

% configurations

$\tilde{K}^\text{bs}$ & local configuration before solidification \\
$\hat{K}$ & plastic intermediate configuration \\
$\tilde{K}^\text{lf}$ & load-free reference configuration\\
$\tilde{K}^\star$ & load-free reference configuration with cancelled volumetric changes \\
$\tilde{K}^\text{sf}$ & stress-free reference configuration\\
$K$ & current configuration \\

%\hline
\end{tabular}
\end{table}

%_____________________________________________________________________

\section{Introduction}

Welded metal structures appear in many applications, including automotive, aerospace, pressure vessels and pipelines.
The  welding involves large thermal gradients due to local heating and cooling, causing structural changes and residual stresses in the end product.
The residual stresses have a significant effect on the  load bearing capacity and fatigue life of welded structures \cite{McCLUNG2006, James2007}.
Moreover, residual stresses in bulk blanks affect milling operations since they lead to undesired deformations of parts.

A conventional way to predict the welding-related residual stresses is to simulate the welding process.
Despite many discretisation methods, finite element method (FEM) is chosen for
analysis of pre-stressed structures \cite{Brickstad1998, Chang2004, Ahn2017, Rae2017}.
Diversified modelling strategies are available; they differ by the set of physical phenomena considered.
Applied theoretical frameworks include structural and thermal analysis \cite{Xie2015, Ahn2017, Yan2018}, simulation of phase transformation \cite{Ahn2017}, volumetric change \cite{Xie2015, Ahn2017, Yan2018}, strain hardening \cite{Wohlfahrt2012, Xie2015, Ahn2017} and creep \cite{Yan2018}.
A proper simulation of each of these phenomena is related to substantial difficulties and ambiguities.
Therefore, as reported by many research groups \cite{Wohlfahrt2012, Ahn2017, Rae2017}, the prediction of residual stresses can be very inaccurate.
High inconsistency of simulation results is caused by a number of interacting aberrations, including ill-defined boundary conditions and complex micro-structural evolutions \cite{Ramirez2003, Ahn2017, Petkov2019}.

After the field of residual stresses is estimated, this field can be used for further analysis in a number of ways.
For instance, the entire pre-stressed structure can be simulated using so-called ``inherent strain'' method \cite{Lu2020} or $\mathbf{F}_0$-approach \cite{Tagiltsev2020}.
This study presents a continued development of the mentioned $\mathbf{F}_0$-approach.

 %Therefore, all the simulations in the current study are carried out using FEM. However, alternative discretization methods like smoothed particle hydrodynamics is also applicable.

In the current paper, elasto-viscoplastic material model with multiplicative split of the deformation gradient \cite{SimoMiehe1992} is used due to its essential advantages: multiplicative framework allows us to build constitutive equations thermodynamically consistent, objective under superimposed rigid body motion, and free from parasitic shear stress oscillations \cite{ShutovIhlemann2014}.
The alternative approach to finite strain kinematics of elasto-plastic material, based on the additive decomposition of the strain-rate tensor, is shown to provide non-physical results for pre-stressed structures \cite{JiaoFish2017}.
For greater accuracy of simulation, various types of hyperelastic potentials can be used within the multiplicaitve approach.
One of classical models of multiplicative elasto-visco-plasticity is the model of Simo and Miehe \cite{SimoMiehe1992}, employed in the current study.
An important symmetry property that constitutive equations remain the same under any isochoric change of reference configuration is called weak invariance (w-invariance): We show that w-invariance is beneficial for the analysis of pre-stressed welded structures.
A significant aspect of the w-invariance is the transformations of initial conditions for plastic strains (see \cite{ShutovIhlemann2014} for a general discussion of w-invariance).
Another important implication of the w-invariance is that it allows us to build efficient computational algorithms \cite{Shutov2016, Shutov2017}.

As mentioned, the theoretical predictions of residual stresses are usually inaccurate \cite{Wohlfahrt2012, Rae2017}.
To improve the accuracy of theoretical computations, we use a method of analyzing stresses that includes a transformation between reference configurations.
Since the transformation is carried out by $\mathbf{F}_0$-mapping, we call this method $\mathbf{F}_0$-approach (cf. \cite{Tagiltsev2020}).
In the current paper it is shown that $\mathbf{F}_0$-approach suits models of multiplicative elasto-viscoplasticity exhibiting w-invariance.
In particular, numerical algorithms for structures with residual stresses are as simple as original algorithms.

Experimental methods of residual stress detection are naturally divided into two groups: destructive and non-destructive.
X-ray and neutron diffraction are conventionally used for non-destructive measurement of residual stresses on the surface of materials \cite{Chang2004, Ahn2017, Rae2017, Yan2018, Lu2020}.
Internal residual stresses are usually measured with destructive methods, e.g. by hole-drilling method and its variations \cite{Smith2000, Peric2019}, strain gauge method \cite{Deng2006, Deng2008}, slitting \cite{Kapadia2017} or contour method \cite{Xie2015}.
Of course, these methods are not acceptable for analysis of unique or expensive structures.
Similar to theoretical predictions discussed priorly, experimental methods exhibit low accuracy, especially evaluating stresses inside a bulky structure
 \cite{Wohlfahrt2012}.

The paper presents hybrid experimental/theoretical method, combining strong points of experimental and numerical techniques.
The advantages of the hybrid method are the following:
stress fields obtained with the $\mathbf{F}_0$-approach satisfy the equilibrium equations exactly, compared to experimental methods that do not;
the proposed technique allows to  ``look inside'' the observed volume;
the obtained stress-field is consistent with experimentally observed stresses at the sites where they are available.

The method's idea is that $\mathbf{F}_0$-field is represented as a function depending on a set of parameters; then a least-square optimisation is implemented to calibrate these parameters against the available experimental data.
The simplest version of technique which does not include a full-scale thermal problem is considered.
The practical applicability of the method is demonstrated on real data for the titanium alloy Ti6Al4V.

The paper is organised as follows.
In Section 2, we formulate the well-known material model of finite-strain elasto-visco-plasticity.
Section 3 has the discussion of the transformation between stress-free and load-free reference configurations ($\mathbf{F}_0$-approach).
Section 4 presents the hybrid experimental/theoretical method for prediction of residual stresses in welds. Demonstration problems are solved, including
the reduction of residual stresses by mechanical treatment. Section 5 contains the discussion of the main results.

%_____________________________________________________________________

\section{Material model of finite-strain elasto-viscoplasticity}

\subsection{Constitutive equations}

The mechanical response is described by the model proposed by Simo and Miehe (1992) \cite{SimoMiehe1992}. We follow its representation on the reference configuration (cf. \cite{Lion1997}).
We consider a system of constitutive equations,  \emph{not a complete model
of finite strain thermoplasticity}.
Although the model is simple,
it accounts for the temperature-induced volume changes
and temperature dependence of most essential material constants.
Reader, interested in consistent approach to thermoplasticity within the multiplicaitve decomposition
of the deformation gradient, is referred to \cite{SimoMiehe1992}, \cite{Lion2000} and \cite{ShutovIhlemannThermoplasticity}, among others.

For certain material point, let $\mathbf{F}$ be the deformation gradient. We denote by
$\mathbf{C} := \mathbf{F}^{\text{T}}\mathbf{F}$ the right Cauchy-Green tensor.
Next, assume the classical multiplicative decomposition of the deformation gradient into the elastic part $\mathbf{F}_{\text{e}}$
and the plastic part $\mathbf{F}_{\text{p}}$
\begin{equation}\label{multiplicative}
\mathbf{F} = \mathbf{F}_{\text{e}} \ \mathbf{F}_{\text{p}}.
\end{equation}
This decomposition gives rise to the
plastic right Cauchy-Green tensor $\mathbf{C}_{\text{p}} := \mathbf{F}_{\text{p}}^{\text{T}}  \mathbf{F}_{\text{p}}$.
To govern the isotropic hardening, we use the plastic arc-length
(Odqvist parameter) $s$, a monotonically increasing
quantity.

Here, a model with linear isotropic hardening is considered. The temperature and
the Helmholz free energy per unit mass are denoted by $\theta$ and $\psi$, respectively.
Let $\psi_{\text{el}}(\mathbf{C}\mathbf{C}_{\text{p}}^{-1})$ and $\psi_{\text{iso}}(s)$ be the energy stored due to macroscopic elastic strains and
the energy stored in crystal lattice defects due to isotropic hardening, respectively.
The free energy function $\psi_{\text{el}}(\mathbf{A})$ is isotropic.
We postulate that the mechanisms of energy storage pertaining to elasticity and isotropic hardening are independent:
\begin{equation}
\psi = \psi_{\text{el}}(\mathbf{C}\mathbf{C}_{\text{p}}^{-1}, \theta) +  \psi_{\text{iso}}(s, \theta).
\end{equation}
Let $\rho_{\text{R}}$ be the mass density in the reference configuration. The
volumetric part of the hyperelastic potential is given by the assumption of Hartmann and Neff \cite{HartmannNeff};
the isochoric part corresponds to the neo-Hookean strain energy:

\begin{equation}\label{ansatz}
\rho_{\text{R}}\psi_{\text{el}}(\mathbf{A}, \theta) =
\frac{k(\theta)}{50}\Bigg( \Big( \frac{\sqrt{ \det \mathbf{A} }}{J_0 (\theta)}
\Big)^{5} + \Big(\frac{\sqrt{ \det \mathbf{A} }}{J_0 (\theta)}\Big)^{-5} -2 \Bigg) +\frac{\mu(\theta)}{2}(\text{tr}\overline{\mathbf{A}} - 3).
\end{equation}
Here, $k(\theta)$ and $\mu(\theta)$ are temperature-dependent bulk and shear moduli;
$J_0 (\theta)$ is the temperature-dependent volume change; $\overline{\mathbf{A}} = (\det\mathbf{A})^{-1/3} \mathbf{A}$
is the unimodular part of a tensor.
The energy storage due to isotropic hardening is assumed as
\begin{equation}\label{quadraticIso}
\rho_{\text{R}}\psi_{\text{iso}}(s, \theta) = \frac{\gamma(\theta)}{2} s^2,
\end{equation}
where $\gamma(\theta)$ is the hardening modulus.

According to the Coleman-Noll procedure, the second Piola-Kirchhoff stress $\widetilde{\mathbf{T}}$ is computed as follows:
\begin{equation}\label{2ndPK}
\widetilde{\mathbf{T}} = 2\rho_{\text{R}} \frac{\partial\psi_{\text{el}}(\mathbf{C}\mathbf{C}_{\text{p}}^{-1})}
{\partial \mathbf{C}}|_{\mathbf{C}_{\text{p}} = const}.
\end{equation}
Using the concrete ansatz \eqref{ansatz} and assuming plastic incompressibility ($\det(\mathbf{C}_{\text{p}}) =  1$), we have
\begin{equation}\label{2ndPKconcrete}
\widetilde{\mathbf{T}} =  \frac{k(\theta)}{10} \Bigg( \Big(\frac{\sqrt{ \det \mathbf{C} }}{J_0 (\theta)}\Big)^5
- \Big(\frac{\sqrt{ \det \mathbf{C} }}{J_0 (\theta)}\Big)^{-5} \Bigg) \mathbf{C}^{-1} +
\mu(\theta) \mathbf{C}^{-1} (\overline{\mathbf{C}} \mathbf{C}_{\text{p}}^{-1})^{\text{D}}.
\end{equation}
Note that the stress tensor's hydrostatic component equals zero whenever $ \det \mathbf{F} = J_0(\theta)$.
In other words, the evolution of $J_0(\theta)$ controls the volume of the stress-free configuration of the material.

The stress-like quantity $R \in \mathbb{R}$, called isotropic hardening, is responsible for isotropic expansion of the yield surface in the stress space.
It is computed as
\begin{equation}
R = \rho_{\text{R}} \frac{\partial\psi_{\text{iso}}(s)}{\partial s} = \gamma(\theta) s.
\end{equation}
A model of viscoplastic type is considered here; stress states beyond the elastic domain are admissible.
The viscous overstress $f$ and the norm $\mathfrak{F}$ of the deviatoric part of the Kirchhoff stress
 are computed using
\begin{equation}
f := \mathfrak{F} - \sqrt{\frac{2}{3}}(K(\theta)+R), \quad
\mathfrak{F} := \sqrt{ \text{tr}[(\mathbf{C} \widetilde{\mathbf{T} } )^{\text{D}}]^2 }.
\end{equation}
Here, $K(\theta)$ is the uniaxial quasi-static yield stress.
The plastic strain rate $\lambda_{\text{p}}$ equals the norm of the inelastic strain rate tensor. It is estimated via the Perzyna law of viscoplasticity:
\begin{equation}
\lambda_{\text{p}} = \frac{1}{\eta}\Big\langle\frac{f}{k_0}\Big\rangle^m, \quad \langle x \rangle := \max(x,0).
\label{Perz}
\end{equation}
Here, $\eta$ and $m$ are the fixed viscosity and stress exponent (temperature-dependence is ignored for simplicity);
$k_0 = 1$ MPa. The temporal change of the internal variables is described by the evolution equations
\begin{equation}\label{flowrule}
\dot{\mathbf{C}}_{\text{p}} = 2\frac{\lambda_{\text{p}}}{\mathfrak{F}}(\mathbf{C}\widetilde{\mathbf{T}})^{\text{D}}\mathbf{C}_{\text{p}}, \quad \dot{s} = \sqrt{\frac{2}{3}}\lambda_{\text{p}}.
\end{equation}
Here, $\mathbf{A}^{\text{D}} = \mathbf{A} - \frac{1}{3} \text{tr}(\mathbf{A}) \mathbf{1}$ is the
deviatoric part of a tensor.
For what follows it is important to note that the plastic flow $\eqref{flowrule}_1$ is indeed incompressible:
\begin{equation}
\det(\mathbf{C}_{\text{p}}) =  1.
\end{equation}
The system of constitutive equations is closed by specifying initial conditions
\begin{equation}\label{InitialConditions}
\mathbf{C}_{\text{p}}|_{t=0} =  \mathbf{C}^0_{\text{p}}, \quad
s|_{t=0} = s^0.
\end{equation}

Note that the model is geometrically exact since it is not based on any geometric linearizations.
Robust and efficient numerical algorithms are available for this model \cite{Shutov2016}.
The model exhibits a set of useful properties:
\begin{itemize}
  \item Thermodynamically consistent in isothermal case: the Clausius-Duhem inequality holds true for arbitrary loading processes \cite{Lion1997}.
  \item Objective (under superimposed rigid body motion).
  \item Free from non-physical shear stress oscillations.
   \item Weakly-invariant under arbitrary isochoric changes of the reference configuration (see next subsection).
\end{itemize}

\textbf{Remark 1.} Since the elastic strains in metals are small, the specific choice of the isotropic storage energy function
is not important. For simplicity, the neo-Hookean potential is utilized here. The reader interested in
a more general anisotropic approach is referred to \cite{Eidel2003}. The use of the spline-based
strain energy functions is discussed in \cite{LatorreMontans2015}.
Efficient numerics for the Mooney-Rivlin elastic potential is described in \cite{Shutov2017}.

\subsection{Weak invariance of the material model under change of the reference configuration}

The material model presented in the previous subsection is local:
In isothermal case, the true stress (Cauchy stress) $\mathbf{T}$ is a unique functional of the local history of the
deformation gradient $\{\mathbf{F}(\tau)\}_{0 \leq \tau \leq t}$ and initial conditions \eqref{InitialConditions}:
\begin{equation}\label{CompactForm}
\mathbf{T}(t) = \mathbf{T} \big(\{\mathbf{F}(\tau)\}_{0 \leq \tau \leq t}, \mathbf{C}^0_{\text{p}}, s^0 \big).
\end{equation}
Note that this relation holds for a chosen fixed reference configuration.
With respect to a new reference configuration, the deformation process
is described by the deformation gradient:
\begin{equation}\label{ReferenceChange}
\mathbf{F}^{\text{new}}(t) = \mathbf{F}(t) \ \mathbf{F}^{-1}_\text{ref.change},
\end{equation}
where $\mathbf{F}_\text{ref.change}$ is the mapping between reference configurations. In this study,
we restrict ourselves to isochoric reference change: $\det \mathbf{F}_\text{ref.change} = 1$.
The following property was proved in \cite{ShutovPfeiffer} in case $J_0 = 1$.
By the transition to the new reference configuration, the material model
predicts the same true stresses, if the initial conditions are transformed
according to the rule ${\mathbf{C}^0_{\text{p}}}^{\text{new}} = \mathbf{F}^{-\text{T}}_\text{ref.change} \ \mathbf{C}^0_{\text{p}} \ \mathbf{F}^{-1}_\text{ref.change}$:
\begin{equation}\label{WeakInvariance}
\mathbf{T} \big(\{\mathbf{F}(\tau)\}_{0 \leq \tau \leq t},
\mathbf{C}^0_{\text{p}}, s^0\big) = \mathbf{T} \big(\{\mathbf{F}(\tau)
\mathbf{F}^{-1}_\text{ref.change}\}_{0 \leq \tau \leq t}, {\mathbf{C}^0_{\text{p}}}^{\text{new}}, s^0\big).
\end{equation}
In other words, only the initial conditions are transformed
upon the change of the reference configuration and the remainig constitutive equations remains the same.
Such a property is called weak invariance (w-invariance), cf. \cite{ShutovIhlemann2014}.

\subsection{Temperature dependence of titanium alloy Ti6Al4V}

For demonstration purposes, we consider the alloy Ti6Al4V; the elasto-visco-plastic model from Section 2.1 is the basis for the following simulations.
The temperature dependence of parameters is described in the following way.
As a reference temperature we chose $\tilde{\theta} = 20 \ {}^{\circ}\text{C}$.
Relative to this state, the non-dimensional volume $Vol$ of the material is described by (cf. \cite{Mills2002, Xie2015})
\begin{equation}
Vol (\theta) = \begin{cases}
    \begin{array}{l}
        1 + 10^{-5}(0.85(\theta-20) + 10^{-4} (\theta-20)^2), \ \text{if }\theta < 770 \ {}^{\circ}\text{C},\\
        1 + 10^{-5}(693.75 + \theta - 20 - 750),\ \text{if }\theta \geq 770 \ {}^{\circ}\text{C}.
    \end{array}
\end{cases}
\end{equation}
Let $\theta_0$ be the initial temperature at the material point.
The non-dimensional volume change relative to the initial state is the ratio
$J_0 (\theta) = Vol (\theta) / Vol (\theta_0)$.
Next, the dependence of the elasticity modulus on the temperature is given by (cf. \cite{Xie2015})
\begin{equation}
E (\theta) = \begin{cases}
    \begin{array}{l}
        1.2 \cdot 10^{5} - (1.2\cdot10^{5} - E_{1300})\theta/1300,\ \text{if }\theta < 1300 \ {}^{\circ}\text{C},\\
        E_{1300} - (E_{1300} - 100)(\theta - 1300)/1200,\ \text{if }\theta \geq 1300 \ {}^{\circ}\text{C},
    \end{array}
\end{cases}
\end{equation}
where $E_{1300}$ is the elasticity modulus at 1300 ${}^{\circ} \text{C}$; the dimension is MPa.
Poisson's ratio equals (cf. \cite{Ahn2017})
\begin{equation}
\nu (\theta) = \begin{cases}
    \begin{array}{l}
        0.34 + 0.06 \ \theta / 1700,\ \text{if }\theta < 1700\ {}^{\circ}\text{C},\\
        0.4 + 0.09(\theta - 1700)/200,\ \text{if } 1700\ {}^{\circ}\text{C} \leq \theta < 1900\ {}^{\circ}\text{C},\\
        0.49,\ \text{if }\theta \geq 1900\ {}^{\circ}\text{C}.
    \end{array}
\end{cases}
\end{equation}
Note that the material is nearly incompressible for $\theta \geq 1900\ {}^{\circ}\text{C}$.
For the use within the material model it is instructive to compute the bulk and shear moduli. The standard relations are
\begin{equation}
k(\theta) = E(\theta) / ( 3 - 6\nu(\theta) ), \quad \mu(\theta) =  E(\theta)/ (2 + 2\nu(\theta)).
\end{equation}
The temperature-dependent yield stress (in MPa) is given by (cf. \cite{Xie2015})
\begin{equation}
K(\theta) =  \begin{cases}
    \begin{array}{l}
        850 - 0.8 \theta,\ \text{if }\theta < 1000\ {}^{\circ}\text{C},\\
        50,\ \text{if }\theta \geq 1000\ {}^{\circ}\text{C}.
    \end{array}
\end{cases}
\end{equation}
The isotropic hardening is assumed to be proportional to the yield stress:
\begin{equation}
\gamma(\theta) = \gamma_{20} K(\theta)/K(20),
\end{equation}
where $\gamma_{20}$ is the isotropic hardening at $20\ {}^\circ \text{C}$.

%_____________________________________________________________________
\section{Transformation of configurations}

\subsection{F0-approach for analysis of residual stresses}

Let us consider a particle in the unloaded body.
A neighbourhood of the particle is associated with a load-free configuration $\tilde{K}^\text{lf}$.
During further deformation, this configuration is transformed to the current configuration $K$ by the deformation gradient $\mathbf{F}^\text{lf}$.
Likewise, let $\tilde{K}^\text{sf}$ and $\mathbf{F}^\text{sf}$ be the local configuration occupied by the particle in the stress-free state and the corresponding deformation gradient.
The difference between configurations is caused by the presence of residual stresses: the residual stresses deform the configuration $\tilde{K}^\text{sf}$ to $\tilde{K}^\text{lf}$.
Conversely, local configuration of the particle transforms from $\tilde{K}^\text{lf}$ to $\tilde{K}^\text{sf}$ during its imaginary local unloading (internal variables like plastic strains are frozen during this imaginary unloading); the deformation gradient related to unloading is denoted as $\mathbf{F}_0$.
Thus, we have
\begin{equation} \label{DefGradsflf}
	\mathbf{F}^\text{sf} = \mathbf{F}^\text{lf} \mathbf{F}^{-1}_0.
\end{equation}
For every particle of the considered structure, deformation gradient $\mathbf{F}_0$ is unique up to superimposed rigid-body rotation.
Since load-free and stress-free configurations are fixed during continued deformation of the body, we can assume $\mathbf{F}_0$ to be constant in time.
Using \eqref{DefGradsflf}, one can equally describe the kinematics of the material particle respective to two different reference configurations.

Observe that the $\mathbf{F}_0$-field is incompatible.
Namely, there is no a displacement field of the entire body corresponding to $\mathbf{F}_0$.
Note that the $\mathbf{F}_0$-field can not be chosen arbitrarily, since the corresponding stress field must satisfy the equilibrium equations.
In the folllowing subsection we discuss one approach to practical identification of the $\mathbf{F}_0$-field.

\subsection{Simulation of solidification}

During welding, temperature of material changes significantly, leading to local distortions and volume changes.
This subsection clarifies the kinematics of the process.

\begin{figure}\centering
    \psfrag{Klf}[m][][1][0]{$\tilde{K}^\text{lf}$}
    \psfrag{Knew}[m][][1][0]{$\tilde{K}^\star$}
    \psfrag{Kbs}[m][][1][0]{$\tilde{K}^\text{bs}$}
    \psfrag{Khat}[m][][1][0]{$\hat{K}(t_\text{solid})$}
    \psfrag{Ksf}[m][][1][0]{$\tilde{K}^\text{sf}$}
    \psfrag{F0}[m][][1][0]{$\mathbf{F}_0$}
    \psfrag{Fbs}[m][][1][0]{$\mathbf{F}^\text{bs}(t_\text{solid})$}
    \psfrag{Fs}[m][][1][0]{$\mathbf{F}^\text{s}$}
    \psfrag{Phi}[m][][1][0]{$\varphi \mathbf{1}$}
    \psfrag{J0}[m][][1][0]{$J_\text{s}^{1/3} \mathbf{1}$}
    \psfrag{Fpbs}[m][][1][0]{$\mathbf{F}_\text{p}^\text{bs}(t_\text{solid})$}
    %Legend psfrags
    \psfrag{L1}[m][][1][0]{$\tilde{K}^\text{bs}$}
    \psfrag{L2}[m][][1][0]{$\tilde{K}^\text{lf}$}
    \psfrag{L3}[m][][1][0]{$\tilde{K}^\star$}
    \psfrag{L4}[m][][1][0]{$\hat{K}(t_\text{solid})$}
    \psfrag{L5}[m][][1][0]{$\tilde{K}^\text{sf}$}
    \psfrag{L6}[m][][1][0]{$\tilde{K}^\text{lf}$}
    \psfrag{L7}[m][][1][0]{$t = t_\text{solid}$}
    \scalebox{1.0}{
    \includegraphics{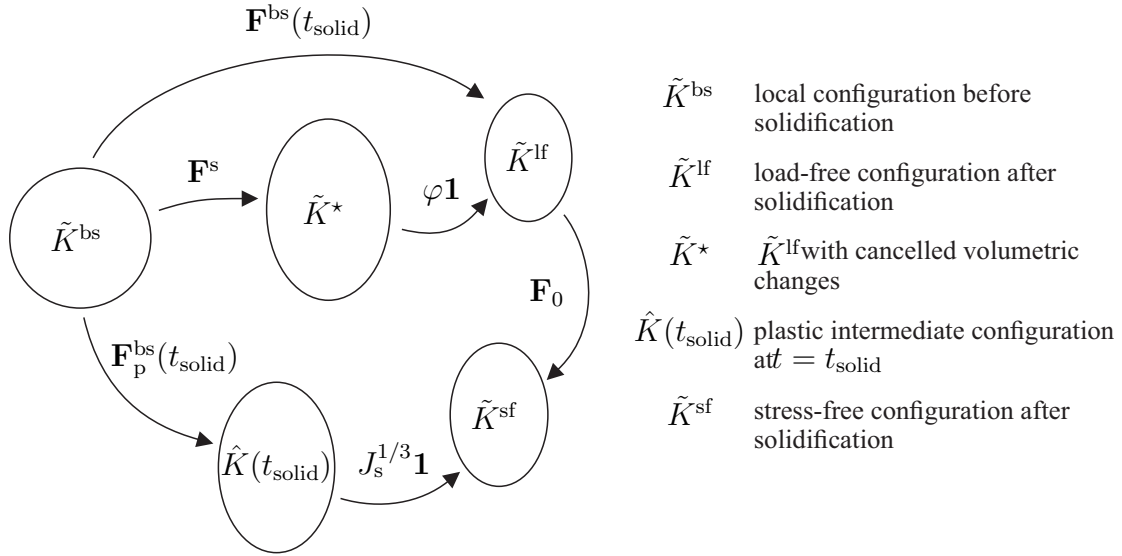}}
    \caption{Commutative diagram showing the kinematics of weld solidification. \label{Scheme1}}
\end{figure}

Let $\tilde{K}^\text{bs}$ be the initial stress-free configuration of a particle before the filler's solidification; let $t_{\text{solid}}$ be the time instance when the solidification is complete. The deformation gradient $\mathbf{F}^\text{bs}$ at the moment $t_\text{solid}$ transforms $\tilde{K}^\text{bs}$ to the load-free configuration $\tilde{K}^\text{lf}$ of this particle after the weld is cooled.
The Flori split of the deformation gradient into isochoric and deviatoric parts reads $\mathbf{F}^\text{bs}(t_\text{solid}) = \varphi \mathbf{1} \cdot \mathbf{F}^\text{s}$; it allows us to introduce new configuration $\tilde{K}^\star$ which corresponds to the deformed state with canceled volume changes (Fig. \ref{Scheme1}).
Here, we use notation $\varphi = (\det \mathbf{F}^\text{bs}(t_\text{solid}) )^{1/3} $ and $\mathbf{F}^\text{s} = \overline{\mathbf{F}^\text{bs}(t_\text{solid})}$.
Meanwhile, we can introduce stress-free configuration $\tilde{K}^\text{sf}$ of the particle in the cooled body as in the previous subsection.
The transformation between the reference configurations $\tilde{K}^\text{bs}$ and $\tilde{K}^\text{sf}$ can be described as a two-step process: isochoric plastic deformation $\mathbf{F}_\text{p}^\text{bs}(t_\text{solid})$ to the intermediate configuration $\hat{K}(t_\text{solid})$ and volumetric expansion/contraction $J_\text{s}^{1/3} \mathbf{1}$; here $ J_\text{s} = J_0(\theta(t_\text{solid})) /  J_0(\theta(t_0))$ is the volume change during the transformation from $\tilde{K}^\text{bs}$ to $\tilde{K}^\text{sf}$.
Thus, the following relation holds:
\begin{equation}\label{Transformation}
	\mathbf{F}^\text{bs}(t_\text{solid}) = \varphi \mathbf{F}^\text{s} = J_\text{s}^{1/3} \mathbf{F}_0^{-1} \mathbf{F}_\text{p}^\text{bs}(t_\text{solid}) .
\end{equation}
The corresponding commutative diagram is shown in Figure \ref{Scheme1}.

In the next subsection we show that only $\mathbf{F}_0$, $J_\text{s}$ and $s(t_\text{solid})$ are required to restart the simulation from the new reference configuration, the load-free configuration of the entire welded structure after solidification.

\subsection{Transformation of initial conditions upon the change of reference configuration in a shrinkable material}

Once the solidification is complete and the body is cooled to room temperature, a significant problem is the simulation of the welded structure subjected to operational loads. In such a simulation, the new reference configuration is naturally $\tilde{K}^\text{lf}$.
For $t>t_\text{solid}$ let $K(t)$ be the current configuration occupied by the body, and $\mathbf{F}^\text{lf}(t)$ be deformation gradient from $\tilde{K}^\text{lf}$ to $K(t)$.
The problem is that particle volume has changed and original constitutive equations formulated on $\tilde{K}^\text{bs}$ cannot be used on $\tilde{K}^\text{lf}$.
However, since the presented material model is w-invariant, we use $\tilde{K}^\star$ instead of $\tilde{K}^\text{bs}$.
Upon this isochoric transformation, the constitutive equations  remain the same and we need only to transform initial conditions from $\tilde{K}^\text{bs}$ to $\tilde{K}^\star$.
In addition, since $\tilde{K}^\star$ is mapped to $\tilde{K}^\text{lf}$ by the isotropic tensor $\varphi \mathbf{1}$, the corresponding stress and strain tensors are transformed by a scalar multiplication.

To establish the transformation from $\tilde{K}^\star$ to $\tilde{K}^\text{lf}$, recall that the scalar $\varphi$ is defined as follows
\begin{equation}
	 \varphi = [\text{det} \mathbf{F}^\text{bs}(t_\text{solid})]^{1/3}.
\end{equation}
At the same time, equation \eqref{Transformation} yields
\begin{equation}
	\mathbf{F}_0 = J_\text{s}^{1/3} \varphi^{-1} \mathbf{F}^\text{bs}_\text{p}(t_\text{solid}) {\mathbf{F}^\text{s}}^{-1}, \quad \text{det} \mathbf{F}_0 = J_\text{s} \varphi^{-3}.
\end{equation}
This allows us to calculate $\varphi$ as
\begin{equation}
	\varphi = [ J_\text{s} / \text{det} \mathbf{F}_0]^{1/3}.
\end{equation}

\begin{figure}\centering
    \psfrag{Klf}[m][][1][0]{$\tilde{K}^\text{lf}$}
    \psfrag{Knew}[m][][1][0]{$\tilde{K}^\star$}
    \psfrag{Khat}[m][][1][0]{$\hat{K}(t)$}
    \psfrag{Kcur}[m][][1][0]{$K(t)$}

    \psfrag{Phi}[m][][1][0]{$\varphi \mathbf{1}$}

    \psfrag{Fpn}[m][][1][0]{$\mathbf{F}_\text{p}^\star(t)$}
    \psfrag{Fn}[m][][1][0]{$\mathbf{F}^\star(t)$}
    \psfrag{Fplf}[m][][1][0]{$\mathbf{F}_\text{p}^\text{lf}(t)$}
    \psfrag{Flf}[m][][1][0]{$\mathbf{F}^\text{lf}(t)$}
    \psfrag{Fe}[m][][1][0]{$\mathbf{F}_\text{e}(t)$}
    %Legend psfrags
    \psfrag{L1}[m][][1][0]{$\tilde{K}^\text{lf}$}
    \psfrag{L2}[m][][1][0]{$\tilde{K}^\star$}
    \psfrag{L3}[m][][1][0]{${K}(t)$}
    \psfrag{L4}[m][][1][0]{$\hat{K}(t)$}
    \psfrag{L5}[m][][1][0]{$\tilde{K}^\text{lf}$}
    \psfrag{L6}[m][][1][0]{$t > t_\text{solid}$}
    \psfrag{L7}[m][][1][0]{$t > t_\text{solid}$}
    \scalebox{1.0}{
    \includegraphics{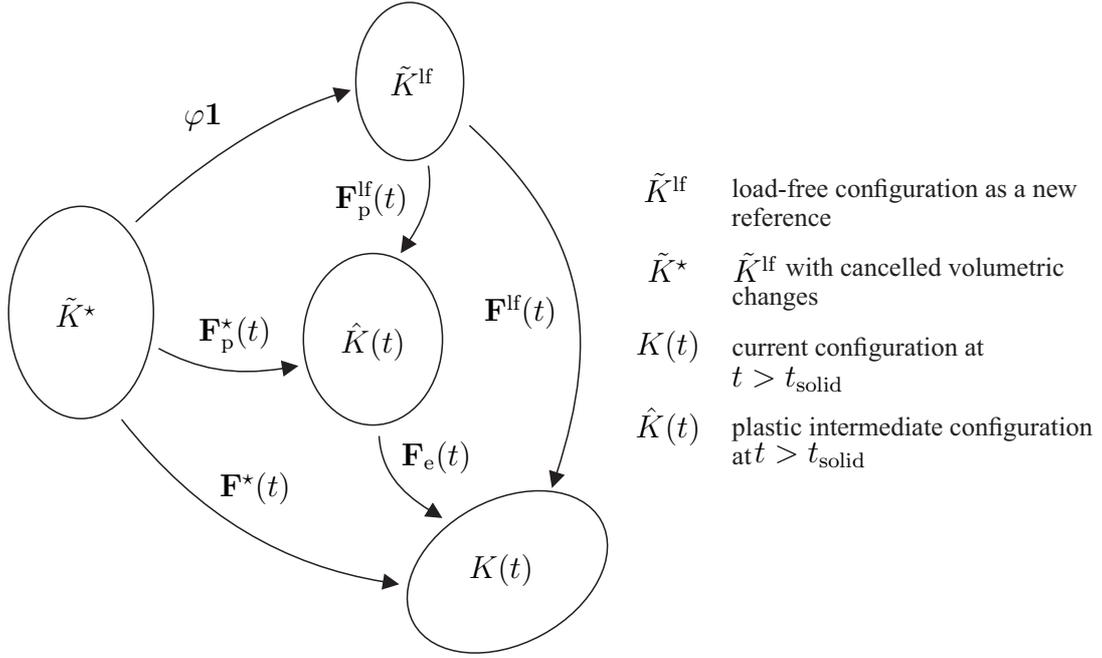}}
    \caption{Commutative diagram for analysis of operational loads occurring after solidification. \label{Scheme2}}
\end{figure}

Introducing the deformation gradient with respect to $\tilde{K}^\star$ as $\mathbf{F}^\star(t) = \varphi \mathbf{F}^\text{lf}(t)$, we derive the following procedure of obtaining second Piola-Kirchhoff stress tensor on $\tilde{K}^\text{lf}$ with use of the original material model \eqref{ansatz} -- \eqref{InitialConditions}.
First, right Cauchy-Green tensor operating on $\tilde{K}^\star$ equals $\mathbf{C}^\star = \varphi^{2} \mathbf{C}^\text{lf}$.
Next, since $\text{det} \mathbf{F}^\text{s} = 1$ and the material model is weakly-invariant, the constitutive equations are used to obtain the second Piola-Kirchhoff stress on $\tilde{K}^\star$: The transformation between $\tilde{K}^\text{bs}$ and $\tilde{K}^\star$ includes the following update of the initial conditions for the plastic right Cauchy-Green tensor:
\begin{equation}\label{PlasticCauchy}
	\mathbf{C}^{\star}_\text{p} \vert_{t_\text{solid}} = (\mathbf{F}^\text{bs}_\text{p}(t_\text{solid}) {\mathbf{F}^\text{s}}^{-1})^\text{T} \mathbf{F}^\text{bs}_\text{p}(t_\text{solid}) {\mathbf{F}^\text{s}}^{-1} = {\mathbf{F}^\text{s}}^\text{-T} \mathbf{C}^\text{bs}_\text{p} {\mathbf{F}^\text{s}}^{-1} = {\overline{\mathbf{F}}}^\text{T}_0 \overline{\mathbf{F}}_0.
\end{equation}
The second Piola-Kirchhoff stress operating on $\tilde{K}^\text{lf}$ and the tangent operator are calculated as
\begin{equation} \label{SimpleFormula}
	\tilde{\mathbf{T}}^\text{lf} = \varphi^{-1} \tilde{ \mathbf{T} }^\star, \quad \frac{\partial \tilde{\mathbf{T}}^\text{lf}}{\partial \mathbf{C}^\text{lf}} = \varphi \frac{\partial \tilde{\mathbf{T}}^\star}{\partial \mathbf{C}^\star}.
\end{equation}
The second relation also holds for consistent tangent operators, required by implicit FEM.
The situation is summarized on the commutative diagram in Figure \ref{Scheme2}.

As seen from \eqref{WeakInvariance}, only the plastic arc-length $s(t_\text{solid})$ and constants $\mathbf{F}_0$ and $J_\text{s}$ are essential to enable restart at $t = t_\text{solid}$ from the load-free configuration, where the initial condition for the plastic strain is given by \eqref{PlasticCauchy}.
Note that the described procedure benefits from the w-invariance of the material model.
In a more general case dealing with models which are not w-invariant one has to introduce the ``pre-processor $\rightarrow$ material model $\rightarrow$ post-processor'' procedure.
Owing to the w-invariance, the entire computational scheme is reduced to computationally efficient scalar multiplication for the right Cauchy-Green tensor, the second Piola-Kirchhoff stress, and the consistent tangent operator.  %during the simulation after restart at $t = t_\text{solid}$.

\textbf{Remark 2.} Observe that not only elasto-plastic models, but also viscoelastic material models based on the multiplicative split and hyperelastic relations between stresses and elastic strains can be made w-invariant.
For instance, popular viscoelastic framework based on the Sidoroff decomposition $\mathbf{F} = \mathbf{F}_\text{e} \mathbf{F}_\text{i}$ can be dealt with likewise.

%_____________________________________________________________________

\section{Residual stresses in the weld}

In order to demonstrate applicability of the advocated method, residual stresses in a Ti-6Al-4V plate after tungsten inert gas (TIG) welding are estimated.

\subsection{Experimental data}

\begin{figure}\centering
    \scalebox{1.0}{
    \includegraphics{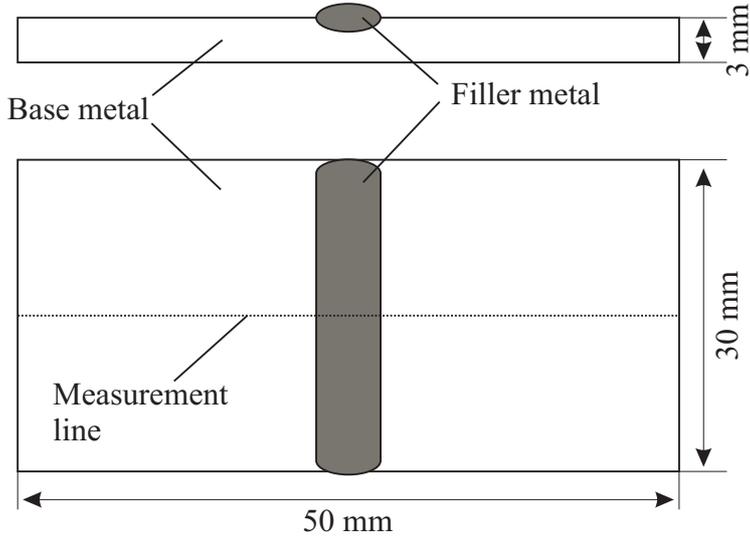}}
    \caption{Schematic illustration of TIG surface welding of a Ti-6Al-4V plate, according to \cite{Yan2018}. \label{ExperimentSetting}}
\end{figure}

The experiment under consideration is described in the work of Yan et al. \cite{Yan2018}.
The TIG surface welding with filler was conducted on Ti-6Al-4V plates; the filler is the same metal.
The dimensions of the plate are 50 mm $\times$ 3 mm $\times$ 30 mm in width, thickness and length, respectively.
The welding was conducted on the top surface of the plate along its middle line (Figure \ref{ExperimentSetting}).
The stresses were measured with X-ray Diffraction (XRD) along the line shown in Fig. \ref{ExperimentSetting}; the total number of 18 measurement points are available; both longitudinal and lateral stresses near the surface were obtained and recorded.
The width of the filler line roughly coincides with the width of fusion zone and equals 6 mm.

\subsection{Parametrisation of the F0-field by FEM simulation}

A simplified problem is considered in FEM analysis, where all the filler material is laid instantly, i.e. the motion of the welding rod is not considered.
This simplifying assumption allows introducing one extra symmetry plane.
Thus, we consider only a quarter of the plate.
Experimental findings from \cite{Yan2018} (see Figure 10c,d in the cited reference) indicate that the residual stresses are nearly symmetric with respect to this extra symmetry plane.
Therefore, the introduction of the symmetry plane is justified.
This assumption allows us to speed up the simulation.

The material model is implemented into the commercial FEM code MSC.MARC via the Hypela2 interface for user-defined models.
An efficient numerical algorithm from \cite{Shutov2016} is used.

\begin{figure}\centering
	\psfrag{Xz}[m][][1][0]{$\{X = 0\}$}
    \psfrag{Zz}[m][][1][0]{$\{Z=0\}$}
    \psfrag{X}[m][][1][0]{$X$}
	\psfrag{Y}[m][][1][0]{$Y$}
	\psfrag{Z}[m][][1][0]{$Z$}
    \scalebox{1.0}{
    \includegraphics{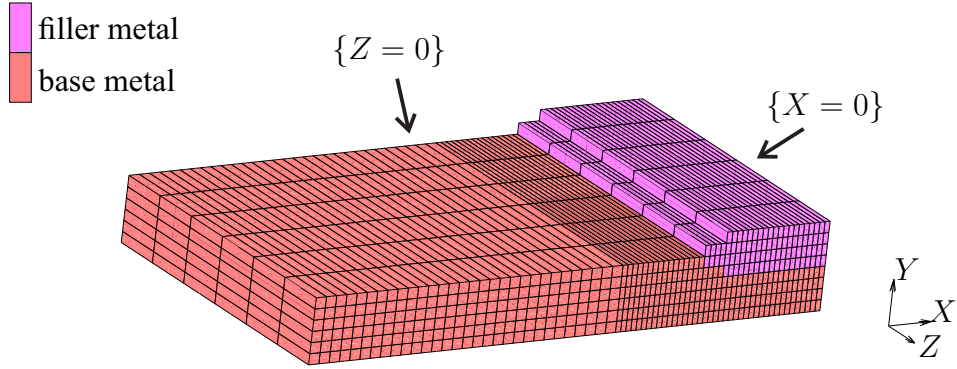}}
    \caption{3D FEM model used for the simulation of welding. \label{ModelFEM}}
\end{figure}

A Cartesian coordinate system is aligned as shown in Figure \ref{ModelFEM}.
The symmetry planes are $\{X = 0\}$ and $\{Z=0\}$; one node on the bottom surface is fixed in Y-direction to prevent undesired rigid body motion.
The boundary of the filler in the XY-plane is an ellipse having the width of 12 mm and height of 2 mm (described as $ X^2/6^2 + (Y-3)^2/1^2 = 1$). The filler is divided by the top horizontal plane $\{Y=3\}$ into two parts of equal volume.

Twenty-node hexagonal elements with a quadratic approximation of the geometry and displacements are used for the modelling.
The size of the mesh is 0.25 $\times$ 0.5 $\times$ 2.5 mm near the filler and 0.5 $\times$ 0.5 $\times$ 2.5 mm for the remaining body.
The total number of elements used is 2784: the filler and the plate are modelled with 528 and 2256 elements, respectively. In the simulation we assume that the plate is made of the base metal.

As already mentioned, the temperature evolution is accounted for in a simple way.
Two different mathematical expressions are used to describe the temperature fields of filler and base metal.
Filler's temperature (in ${}^\circ \text{C}$) starts from its peak and decreases during solidification, the difference between material and environment temperatures halves every 10 seconds \cite{Arunkumar2019}:
\begin{equation} \label{TempFiller}
	\theta_\text{filler}(t) = 20 + 2400 \cdot 2^{-0.1 t} .
\end{equation}
The temperature of base metal changes in two stages.
The first stage is the rapid increase in temperature; it continues for $t_1$ seconds ($t_1 = 3 \text{ s}$ in current paper).
At $t \leq t_1$ the temperature is a linear function of the distance to the filler.
The second stage ($t > t_1$) is the subsequent cooling of the base metal. During this stage, the difference between material and environment temperatures halves every 10 seconds.
Thus, the overall temperature field in the base metal is calculated as following (in ${}^\circ \text{C}$):
\begin{equation} \label{TempBase}
		\begin{cases}
			\theta_\text{base}(t, d) = 20 + \frac{t}{t_1} R(d) \cdot C_\theta \cdot (\theta_\text{filler}(t_1) - 20), \text{ if } t \leq t_1,			
			\\
			\theta_\text{base}(t, d) = 20 + (\theta_\text{base}(t_1, d) - 20)\cdot 2^{-0.1(t - t_1)}, \text{ if } t > t_1
		\end{cases} ,
\end{equation}
where $C_\theta$ is a non-dimensional coefficient defining the dependence between filler's and base's temperatures and
\begin{equation}
	R(d) = \max ( 0 ,  (d_\text{max} - d) / d_\text{max} ).
\end{equation}
Here, $d_\text{max}$ is a maximum expected heating distance (in mm), $d$ is the distance between filler and the current point (in mm).
In the simulation, the solidification time is $t_\text{solid} = 150 \text{ s}$; the temperature of the welded structure at the last step of simulation differs from the environmental temperature $20{}^\circ\text{C}$ by less than $0.1{}^\circ\text{C}$.
Thus, we assume that the cooling is complete at $t=t_\text{solid}$.
Quantities $\mathbf{F}_0$ and $J_\text{s}$ are calculated and saved for every Gauss integration point in that last step of the simulation of solidification.

Quantities $C_\theta$ and $d_\text{max}$ governing the temperature distribution are not known \emph{a priory}.
Moreover, the value of Young's modulus at the temperature $\theta = 1300 {}^\circ \text{C}$ and the isotropic hardening $\gamma_{20}$ are also assumed to be free.
Together with $C_\theta$ and $d_\text{max}$ they will be used to parametrise the $\mathbf{F}_0$-field.

\subsection{Optimisation problem for the estimation of residual stresses}

As  stated in Section 4.1, the residual stresses are measured at 18 points.
Let the measurement points be designated as $\text{MP}_i$ for $i = 1,\ldots,18$.
Measured lateral and longitudinal stresses are denoted as $\mathbf{T}_\text{exp, X}(\text{MP}_i)$ and $\mathbf{T}_\text{exp, Z}(\text{MP}_i)$;
standard deviation of the measurement are denoted as $\text{SD}_\text{X}^{\text{MP}_i}$ or $\text{SD}_\text{Z}^{\text{MP}_i}$.

For the fixed set of modelling parameters, one can obtain numerical values of the true stress at the same points: $\mathbf{T}_\text{num, X}(\text{MP}_i)$ and $\mathbf{T}_\text{num, Z}(\text{MP}_i)$.
Using these, we build the following error functional:
\begin{equation} \label{ErrorFunc}
	\Phi = \sum_{i=1}^{18} (\omega_{i,\text{X}} \cdot [\mathbf{T}_\text{num, X}^{(\text{MP}_i)} - \mathbf{T}_\text{exp, X}^{(\text{MP}_i)}])^2 + \sum_{i=1}^{18} (\omega_{i,\text{Z}} \cdot [\mathbf{T}_\text{num, Z}(\text{MP}_i) - \mathbf{T}_\text{exp, Z}(\text{MP}_i)])^2,
\end{equation}
where $\omega_{i, \text{X}}$ and $\omega_{i, \text{Z}}$ are weighting coefficients, taken as reciprocal values of standard deviations $\text{SD}_\text{X}^{\text{MP}_i}$ and $\text{SD}_\text{Z}^{\text{MP}_i}$ to provide more error-resistant optimisation \cite{Beck1977, ShutovKaygorodtseva2019}.

%Since the points 1, 7, 9, 10, 12 and 18 correspond to the extreme values of measured stresses, weighting coefficients for these points are increased:

%\begin{equation} \label{Weights}
%	\omega_{i, \text{A}} = \begin{cases}
%		3 / \text{SD}_\text{A}(\text{MP}_\text{i}), \text{ if }i \in \lbrace 1, 7, 9, 10, 12, 18 \rbrace
%		\\
%		1 / \text{SD}_\text{A}(\text{MP}_\text{i}), \text{ otherwise}
%	\end{cases},
%	\text{A} \in \lbrace \text{X}, \text{Z}\rbrace.
%\end{equation}

The optimal set of modelling parameters minimizes \eqref{ErrorFunc}. After the best suited parameters are found, they are used to evaluate the $\mathbf{F}_0$-field, $J_\text{s}$ and the residual stresses in the entire structure.
Note that with increasing number of parameters the optimisation problem may become badly posed due to overparametrisation.
Thus, a reasonable amount of free parameters should be used.
Three parameters are subject to identification in this study: the geometry of heat distribution ($d_\text{max}$), the uniformity of heat distribution ($C_\theta$) and a strength parameter. This third parameter can be either $E_{1300}$ or $\gamma_{20}$.
In the following, examples for two parametrisations are considered:

\textbf{Parametrisation with $\text{E}_{1300}$}

\begin{figure}\centering
	\psfrag{C33}[m][][1][0]{$\mathbf{T}_{33},\text{ MPa}$}
	\psfrag{CC}[m][][1][0]{$\mathbf{T}_{11},\text{ MPa}$}
	\psfrag{X1}[m][][1][0]{$X,\text{ mm}$}
	\psfrag{X2}[m][][1][0]{$X,\text{ mm}$}
	\psfrag{E1300}[m][][1][0]{parametrisation}
	\psfrag{E1301}[m][][1][0]{with $E_\text{1300}$}
    \psfrag{G20}[m][][1][0]{parametrisation}
    \psfrag{G21}[m][][1][0]{with $\gamma_\text{20}$}
    \psfrag{Exp1}[m][][1][0]{Experimental}
    \psfrag{Exp2}[m][][1][0]{values}
    \scalebox{1}{
    \includegraphics{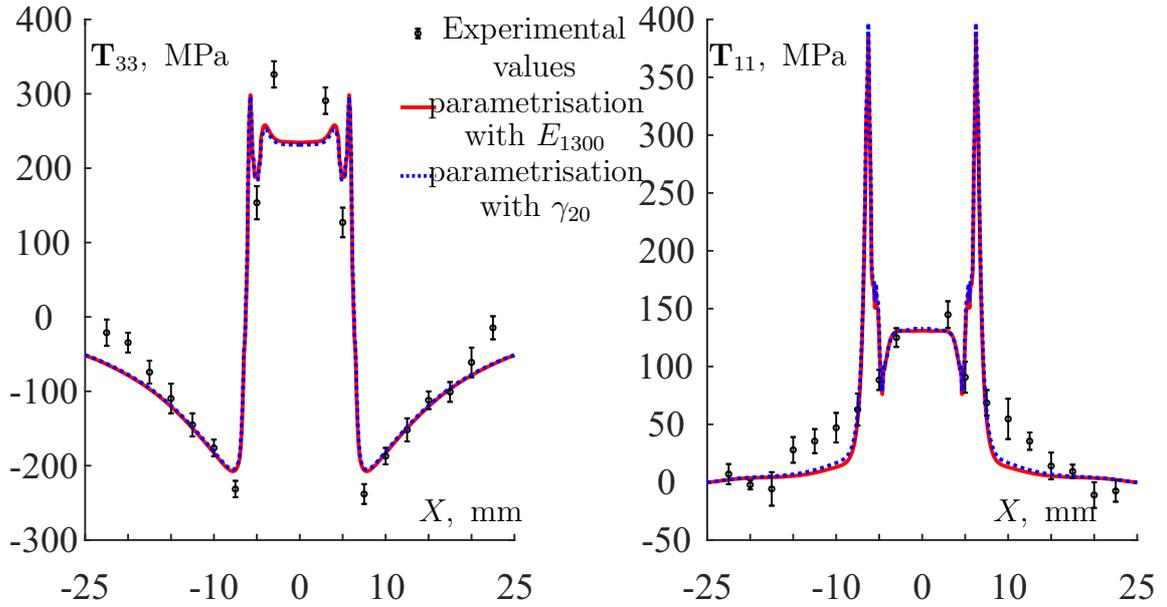}}
    \caption{Stress distribution along the measurement line; experimental values from \cite{Yan2018} are indicated as circles with extra bars corresponding to the errors of measurement, numerical values for the $E_\text{1300}$-parametrisation and $\gamma_\text{20}$-parametrisation are shown with the solid and dashed lines. Left: residual stresses in longitudinal direction. Right: residual stresses in lateral direction. \label{StressOpt}}
\end{figure}

Preliminary numerical studies have shown that the parameter $E_{1300}$ has a clear impact on the residual stress field.
Meanwhile, the experimental identification of this parameter is prone to certain errors, that can influence greatly simulation results.
Therefore, we include this parameter for the parametrisation of the $\mathbf{F}_0$-field.
Thus, the set of parameters used for the first optimisation is $\{ d_\text{max}, C_\theta, E_{1300}\}$, whilst $\gamma_{20}$ is fixed as $5000 \text{ MPa}$.
The optimisation problem is solved with the Nelder-Mead procedure for minimisation of $\Phi$ (cf. \eqref{ErrorFunc}); it is a robust gradient-free method.\footnote{In contrast to gradient-based methods like the Levenberg-Marquardt method, it has a lower probability of getting stuck at a local minimum.}
The initial approximation of parameters is $ d_\text{max}^0 = 25.0 \text{ mm}$, $C_\theta^0 = 0.8$, $E_{1300}^0 = 10000 \text{ MPa}$.
Optimised values are $ d_\text{max} = 30.429 \text{ mm}$, $C_\theta = 0.603$, $E_{1300} = 10640 \text{ MPa}$.
Simulated stresses along the measurement line for this particular parametrisation are shown in Figure \ref{StressOpt} together with available experimental data.

\textbf{Parametrisation with $\gamma_{20}$ }

Similar to the parameter $E_{1300}$, $\gamma_{20}$ substantially impacts the residual stress field (the same observation was made in \cite{Wohlfahrt2012}).
Here we study a strategy based on variation of $\gamma_{20}$.
The set of parameters used for the optimisation is $\{ d_\text{max}, C_\theta, \gamma_{20}\}$, whilst $E_{1300}$ is fixed as $10000 \text{ MPa}$ corresponding to the experimental value \cite{Xie2015}.
The error minimization is carried out with the Nelder-Mead method.
Initial values of parameters for the first iteration are taken as $ d_\text{max}^0 = 25.0 \text{ mm}$, $C_\theta^0 = 0.8$, $\gamma_{20}^0 = 5000 \text{ MPa}$.
Optimised values are $ d_\text{max} = 31.194 \text{ mm}$, $C_\theta = 0.688$, $\gamma_{20} = 5190 \text{ MPa}$.
Computed stresses along the measurement line for this parametrisation are shown in Figure \ref{StressOpt}.

\begin{figure}\centering
	\psfrag{X}[m][][1][0]{$X$}
	\psfrag{Y}[m][][1][0]{$Y$}
	\psfrag{Z}[m][][1][0]{$Z$}
    \scalebox{1}{
    \includegraphics{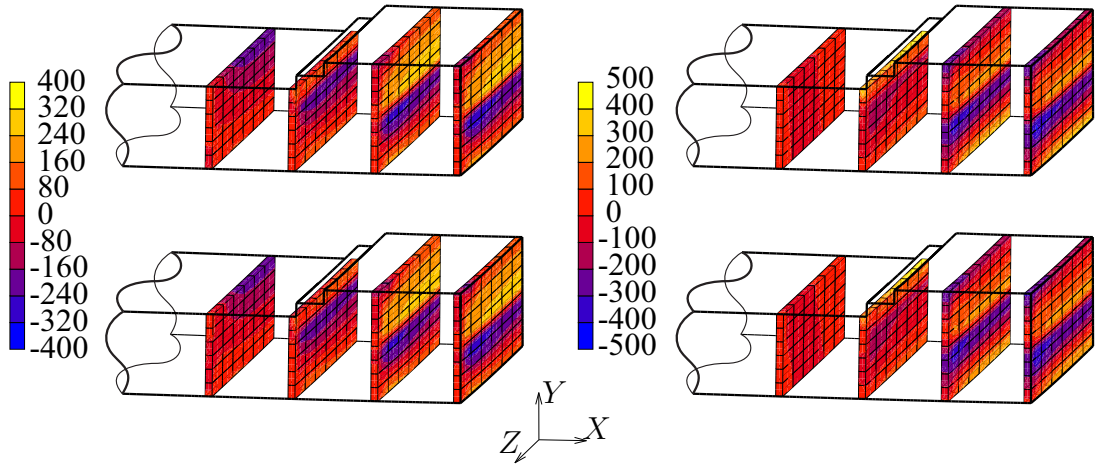}}
    \caption{Residual stress distribution inside the structure; four planes are shown: $\{X=0\text{ mm}\}$, $\{X=-3\text{ mm}\}$, $\{X=-6\text{ mm}\}$ and $\{X=-9\text{ mm}\}$. Top: parametrisation with $E_\text{1300}$. Bottom: parametrisation with $\gamma_\text{20}$. Left: residual stresses in longitudinal direction (in MPa). Right: residual stresses in lateral direction (in MPa). \label{Tomography}}
\end{figure}

As seen, two considered strategies lead to similar results, even though the optimised parameters differ.
Additionaly, the stress distribution inside the structure is visualised in Figure \ref{Tomography}.
Again, the obtained stresses match for both parametrisations.
The results obtained by parametrisation with $E_\text{1300}$ are used to simulate the further loading of the weld after the solidification.

\subsection{Loading of the weld}

Two different cases of subsequent loading are considered: three-point bending and axial stretching.

\textbf{Three-point bending}

\begin{figure}\centering
	\psfrag{X}[m][][1][0]{$X$}
	\psfrag{Y}[m][][1][0]{$Y$}
	\psfrag{Z}[m][][1][0]{$Z$}
    \scalebox{1}{    \includegraphics{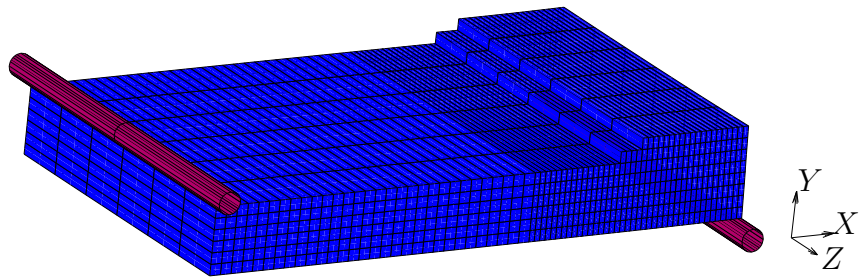}}
    \caption{3D FEM model used for the simulation of three-point bending. \label{ThreePoint3D}}
\end{figure}

\begin{figure}\centering

	\psfrag{C33}[m][][1][0]{$\mathbf{T}_{33},\text{ MPa}$}
	\psfrag{CC}[m][][1][0]{$\mathbf{T}_{11},\text{ MPa}$}
	\psfrag{X1}[m][][1][0]{$X,\text{ mm}$}
	\psfrag{X2}[m][][1][0]{$X,\text{ mm}$}
	\psfrag{Residual}[m][][1][0]{After welding}
    \psfrag{D1}[m][][1][0]{$\text{D}_\text{cyl} = 1.0\text{ mm}$}
    \psfrag{D2}[m][][1][0]{$\text{D}_\text{cyl} = 2.0\text{ mm}$}
    \psfrag{D3}[m][][1][0]{$\text{D}_\text{cyl} = 3.0\text{ mm}$}
    \scalebox{1}{    \includegraphics{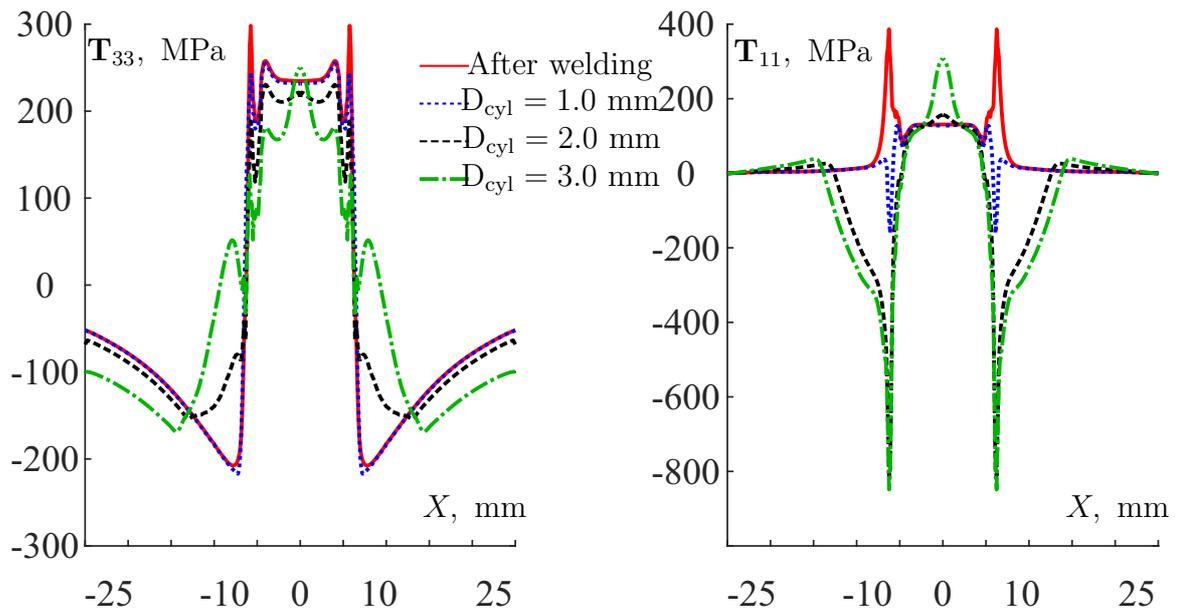}}
    \caption{Stress distribution along the measurement line before and after three-point bending. Left: stresses in longitudinal direction. Right: stresses in lateral direction. \label{ThreePoint}}
\end{figure}

The welded plate is bent by two rigid circular cylinders with diameter of 1 mm; the cylinders are aligned along the Z-direction (\ref{ThreePoint3D}).
One of the cylinders is fixed below the plate under the weld, and
the other is placed above the edge of the plate.
During the bending, the upper cylinder moves along the Y-direction and then it
returns to its initial position. The maximum displacement of the cylinder defines the deformation of the plate.
Three simulations are carried out with maximum displacements of the cylinder $\text{D}_\text{cyl}$ equal to $1.0$, $2.0$, and $3.0$ mm.
Stress distributions along the measurement line before the bending and at the last simulation step are shown in Figure \ref{ThreePoint}.

The interface between the filler and the plate becomes more stressed in the lateral direction after the bending as the stresses are redistributed throughout the plate.
However, it is possible to effectively re-distribute lateral stresses in relatively small bending (with $\text{D}_\text{cyl} = 1.0\text{ mm}$): resulting stresses are considerably smaller than in initial state just after solidification.
Moreover, dangerous tensile stresses are replaced by compressive stresses (Figure \ref{ThreePoint}, right).

\textbf{Stretching along the weld}

The second type of loading is a conventional way of relaxing peak residual stresses \cite{Farajian2013, Lee2015, Wang2017, Schubnell2020}.
The front side of the plate is stretched in the Z-direction followed by unloading.
%The displacement of the front side is prescribed in axial direction. % (every node is linked in Z-direction to a master node, which is displacement-controlled in Z-direction).
%Since the stresses on the measurement line are of interest, Saint-Venant's principle allows to neglect changes in the shape of front side.
%Absence of reaction force in Z-direction indicates reaching the unloaded state of the plate.
Three simulations are carried out with the maximum displacement of the side $\text{D}_\text{front}$ of $0.075$, $0.15$ and $0.225$ mm. This corresponds to $0.5 \%$, $1.0 \%$, and $1.5 \%$ of the plate's length.

\begin{figure}\centering
	\psfrag{C33}[m][][1][0]{$\mathbf{T}_{33},\text{ MPa}$}
	\psfrag{CC}[m][][1][0]{$\mathbf{T}_{11},\text{ MPa}$}
	\psfrag{X1}[m][][1][0]{$X,\text{ mm}$}
	\psfrag{X2}[m][][1][0]{$X,\text{ mm}$}
	\psfrag{Residual}[m][][1][0]{Residual stress}
    \psfrag{D1}[m][][1][0]{$\text{D}_\text{front} = 0.5\%$}
    \psfrag{D2}[m][][1][0]{$\text{D}_\text{front} = 1.0\%$}
    \psfrag{D3}[m][][1][0]{$\text{D}_\text{front} = 1.5\%$}
    \scalebox{1}{
    \includegraphics{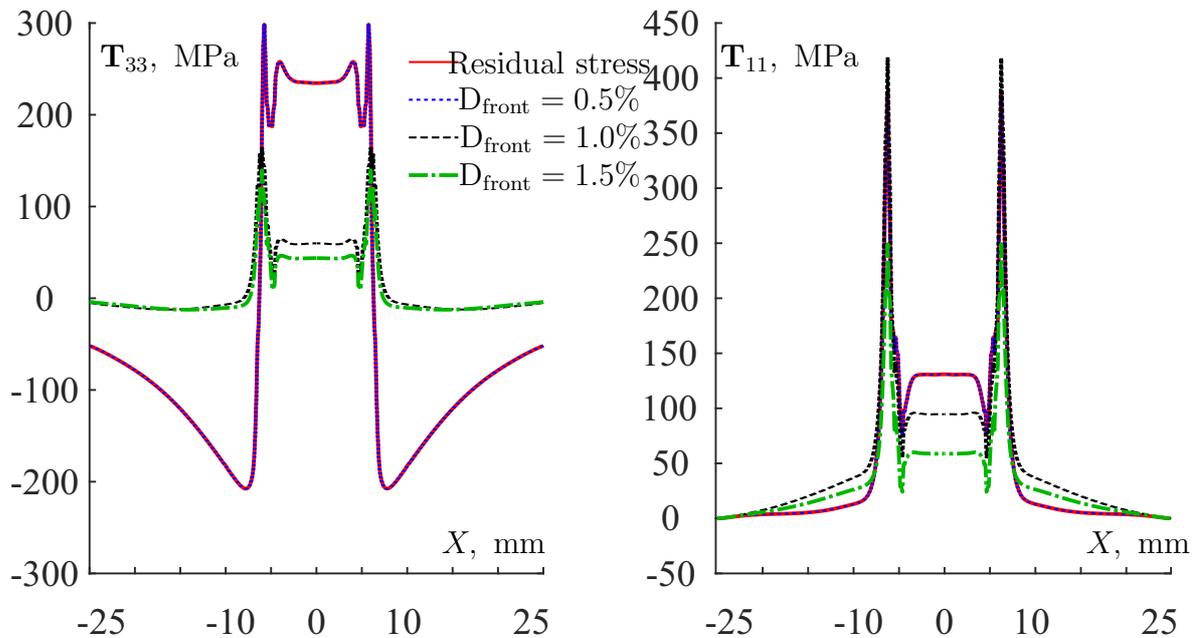}}
    \caption{Stress distribution along the measurement line before and after axial stretching. Left: stresses in longitudinal direction. Right: stresses in lateral direction. \label{ZRelaxation}}
\end{figure}

We see that $0.5 \%$ stretching does influence the residual stresses (Figure \ref{ZRelaxation}).
This effect is due to purely elastic loading below the yield limit.
However, $1.0 \%$ and $1.5 \%$ stretching has a considerable effect.
Longitudinal stresses computed on the measurement line are becoming smaller than in the freshly welded plate.
At the same time, lateral stresses become more uniform.
Similar to
experimental and theoretical results from \cite{Lee2015, Wang2017},
the residual stresses are relaxed by stretch.

\textbf{Comparison of end-to-end and restarted simulations}

\begin{figure}\centering
	\psfrag{C33}[m][][1][0]{$\mathbf{T}_{33},\text{ MPa}$}
	\psfrag{CC}[m][][1][0]{$\mathbf{T}_{11},\text{ MPa}$}
	\psfrag{X1}[m][][1][0]{$X,\text{ mm}$}
	\psfrag{X2}[m][][1][0]{$X,\text{ mm}$}
	\psfrag{Gt1}[m][][1][0]{Go-through}
	\psfrag{Gt2}[m][][1][0]{simulation}
	\psfrag{Gt3}[m][][1][0]{($\tilde{K}^\text{bs}$ as the}
	\psfrag{Gt4}[m][][1][0]{reference)}
	\psfrag{Gt5}[m][][1][0]{ }
	\psfrag{Rs1}[m][][1][0]{Simulation with}
	\psfrag{Rs2}[m][][1][0]{restart after}
	\psfrag{Rs3}[m][][1][0]{solidification}
	\psfrag{Rs4}[m][][1][0]{($\tilde{K}^\text{lf}$ as the new}
	\psfrag{Rs5}[m][][1][0]{reference)}
	\psfrag{Rs6}[m][][1][0]{ }
    \scalebox{1}{
    \includegraphics{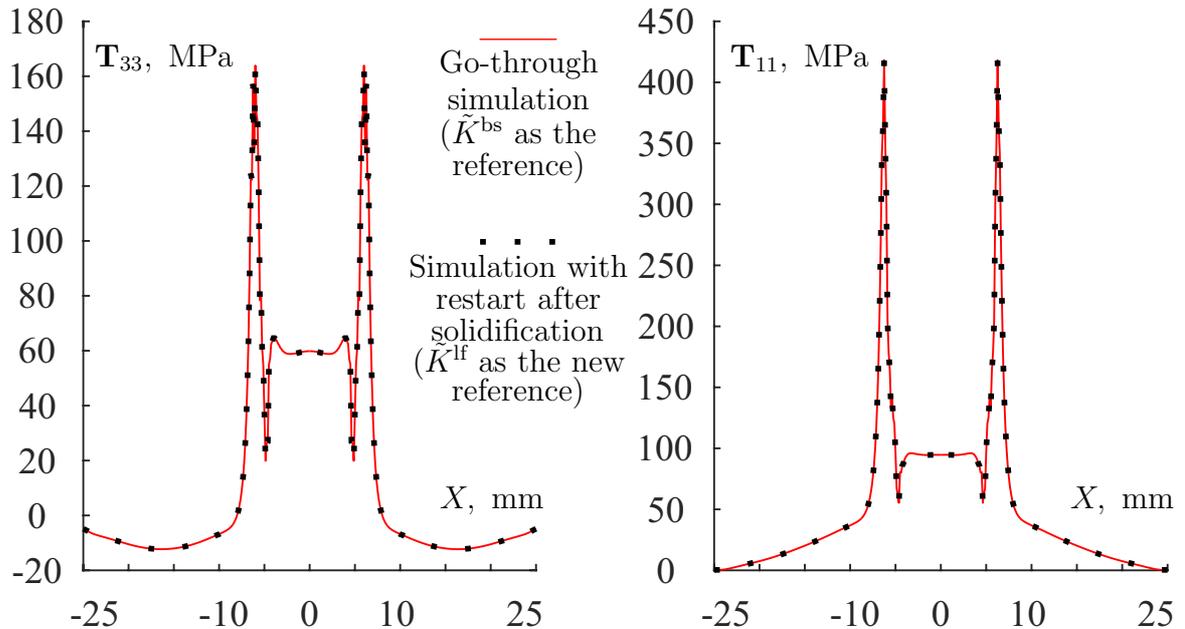}}
    \caption{Stress distributions along the measurement line after axial stretching with $\text{D}_\text{front}~=~1.0\%$. Two simulations are carried out:
    (i) simulation with restart at $t=t_{\text{solid}}$ using the $\mathbf{F}_0$-approach;
    (ii) end-to-end simulation of welding and stretching. Left: stresses in longitudinal direction. Right: stresses in lateral direction.
\label{Compar}}
\end{figure}

To test the theoretical results presented in Section 3.3, to different simulations are carried out.
The first simulation utilizes restart at $t=t_{\text{solid}}$ after weld solidification. After the re-start, the global load-free configuration of the weld $\tilde{K}_\text{lf}$ is the new reference configurations used in MSC.MARC.
The second simulation is the end-to-end simulation without restart; the initial reference $\tilde{K}^\text{bs}$ is used throughout the entire computation in MSC.MARC.
Axial loading with the maximum front side displacement of $0.15 \text{ mm}$ is used in this test.
Computed residual stresses along the measurement line are shown in Figure \ref{Compar} for both simulations.
As is seen, results are identical as expected.

%_____________________________________________________________________

\section{Conclusion and discussion \label{Concl}}

A special $\mathbf{F}_0$-approach to modelling of residual stresses is considered in this work.
In contrast to \cite{Tagiltsev2020}, where $\mathbf{F}_0$-approach was used to establish relationship between $K^\text{lf}$ and $K^\text{sf}$ with $\det \mathbf{F}_0 = 1$, in this paper the approach is generalised to include volumetric changes.

The advocated approach is beneficial for elasto-plastic models exhibiting the weak invariance. In the current study, such a model is based on the multiplicative decomposition of the deformation gradient tensor. Some models based on the additive decomposition of the strain rate tensor are w-invariant as well \cite{ShutovIhlemann2014}.
The positive effect of the w-invariance is that introduction of residual stresses through the $\mathbf{F}_0$-field does not increase the complexity of the numerical schemes.
The stresses and the tangent operator are calculated in the new reference configuration in a straight-forward way using \eqref{SimpleFormula}.
Thus, computationally expensive pre- and post-processors are not required.

Simulations of residual stresses in welds are notoriously oversensitive to errors in constitutive modelling and inaccurate calculation of temperature distribution \cite{Wohlfahrt2012}.
Therefore, it is an impressive result that using a simple material model and a schematic description of temperature evolution, mechanically reasonable results can be obtained:
\begin{itemize}
	\item the method allows to obtain residual stresses in exact equilibrium;
	\item the correspondence between the simulation and the available experimental data (cf. Figure \ref{StressOpt}) is comparable to other simulations based on advanced material models \cite{Yan2018}.
\end{itemize}
In fact, the advocated approach allows extrapolating the experimental data from the surface into the structure.
In the considered example, the extrapolation is stable with respect to chosen parametrisation (Figure \ref{Tomography}).
We emphasize the flexibility of this approach.
In a possible extension, one can carry out a fully-coupled thermo-mechanical simulation of the solidification process and use advanced models of hardening \cite{Shutov2012, Zhang2019, Manopulo2020, Barlat2020}, provided that these models are w-invariant.
In that case, one will be inevitably confronted with poorly-defined boundary conditions and constitutive assumptions;
the critical poorly-defined or most inaccurate simulation parameters can be used for parametrisation of the $\mathbf{F}_0$-field.
Then, the parameters that minimize the error functional of type \eqref{ErrorFunc} estimate the $\mathbf{F}_0$-field and the residual stresses.

The method is promising for mechanobiology and mechanics of composite materials.
The implemented plasticity model can be useful for the prediction of residual stresses caused by chemical curing \cite{Parlevliet2007, Landgraf2018, Jorgensen2019}.
The more accurate simulation of residual stresses allows for a rational analysis of deformation and damage in pre-stressed structures.%
%The solution of the demonstration problem shows that the $\mathbf{F}_0$-approach is a practical tool of analysis. It allows to speed up the numerical optimization of mechanical %treatment used to reduce undesired residual stresses in welds.

\textbf{Acknowledgements.}
The financial support provided by the Russian Science Foundation (project No. 19-19-00126) is acknowledged.

%_____________________________________________________________________

\end{document}